\documentclass[twocolumn,showpacs,preprintnumbers,amsmath,amssymb]{revtex4}
\begin{document}

\title{Violation of local realism vs detection efficiency} 
\author{Serge Massar}
\altaffiliation[Also at]{
Ecole Polytechnique, C.P. 165, 
Universit\'e Libre de Bruxelles, 1050 
Brussels, Belgium}
\author{Stefano Pironio}
\affiliation{Service de Physique Th\'eorique,
Universit\'e Libre de Bruxelles,
 C.P. 225, Bvd. du Triomphe, 1050
Bruxelles, Belgium} 
\date{\today}
\begin{abstract}
We put bounds on the minimum detection efficiency necessary to violate local 
realism in Bell experiments. These bounds depends of simple parameters like the 
number of measurement settings or the dimensionality of the entangled quantum 
state. We derive them by constructing explicit local-hidden variable models 
which reproduce the quantum correlations for sufficiently small detectors 
efficiency. \end{abstract}


\maketitle

\section{Introduction}

Since the work of Bell \cite{Bell} it is well known that "non-local"
correlations can be extracted from entangled states by performing certain 
measurements on spatially separated regions. More precisely
by "non-local" correlations, one
means correlations that cannot be reproduced by local realistic theories. 
Non-locality and  entanglement are closely connected and they form a most 
remarkable features of quantum mechanics. But the relation between these two 
concepts is still not perfectly clear. One difference between them is that while 
entanglement is a characteristic \emph{per se} of a quantum system, 
non-locality depends on the specific experiment carried on the quantum system,
in particular it depends on the measurements performed and on practical details 
such as the efficiency and the background of the detectors, the amount of noise 
present, etc. It is therefore much more difficult to compare non-locality 
exhibited by different experiments and to find measures of non-locality than it 
is for entanglement.

Possible ways to quantify the non-local character of quantum correlations 
exploit their dependence to experimental imperfections like the maximum amount 
of noise or the minimum detection efficiency still allowing a violation of local 
realism. The amount of communication needed to reproduce the quantum 
correlations in a classical scenario can also serve to gauge their non-local 
nature.

In the present paper we concentrate on the resistance to inefficient detectors 
and try to put bounds on how much increase in non-locality can be expected from 
that point of view. We suppose that each detector has a probability $\eta$ of 
giving a result and a probability $1-\eta$ of not giving a result. If $\eta$ is 
sufficiently small, the quantum correlations produced in a Bell experiment can 
be explained by a local hidden variables (LHV) model. We denote by $\eta^*$ the 
maximum detection efficiency for which a LHV model exists. Thus if 
$\eta>\eta^*$ the correlations are indeed non-local. To put bounds on $\eta^*$, 
we construct several LHV models that take advantage of the inefficiency 
of the detectors to reproduce the quantum correlations. LHV models exploiting 
the detection loophole have already been constructed to reproduce the result of 
specific experiments \cite{Santos, Szabo}. There have also been attempts to 
build more general LHV models that can for example reproduce measurements 
performed on the singlet state \cite{Gisin, Larsson} or experiments performed 
using parametric-down conversion sources \cite{cas}. In this paper we try to be 
more general than that. Indeed, our purpose is to understand how $\eta^*$ is 
constrained by simple parameters such as the number of measurements settings or 
the dimensionality of the quantum system. We therefore introduce a first LHV 
model in section \ref{lhvsett} which depend only on the number of measurements 
settings at each site (it is a generalization of a model first discussed in 
\cite{Gisin} and \cite{M}). We describe it both in the case of two parties and 
in the case of many parties. In the case of two measurements per site, the bound 
on $\eta^*$ our LHV model implies is saturated by Eberhard's \cite{Eb} and 
Larsson and Semitecolos's \cite{Larsson2} schemes. In section \ref{lhvdim}, we 
introduce a second model for maximally entangled states that depend only of the 
dimension $d$ of the Hilbert space and which reproduce the quantum correlations 
up to small errors. This LHV model will be analyzed in the case of two parties,
altough it could probably be generalized to more parties. These two LHV models 
work for arbitrary measurements (POVM's) carried out by the parties. Before 
presenting them, let us briefly recall the principle of Bell experiments and the 
content of LHV theories.

\section{Bell experiments and LHV theories}\label{bell}
In a typical Bell experiment, two parties Alice and 
Bob (the generalization to $N$ parties is straightforward) share an entangled 
state $\rho_{AB}$. Alice selects one of $M_A$ measurements on his sub-system and
Bob one of $M_B$. We will consider the most general type of measurements, namely 
Positive Operator Valued Measurements (POVM). Let $X$ be Alice's measurement and 
$Y$ be Bob's measurements and $a$ and $b$ Alice and Bob's outcomes. The POVM $X$ 
thus consists of the positive operators $x_a$ with the property that $\sum_a x_a 
= I_A$. Similarly the POVM $Y$ consists of the positive operators $y_b$ 
with the property that $\sum_b y_b = I_B$. Here $I_A$ and 
$I_B$ are the identity operators. Quantum mechanics predicts the 
probabilities 
\begin{eqnarray}\label{qc} P^{QM}(a,b|X,Y)&=& \mbox{Tr}(x_a 
\otimes y_b \ \rho_{AB})\ , \nonumber \\ 
P^{QM}(a|X)&=& \mbox{Tr}(x_a \otimes I \ 
\rho_{AB})\ , \nonumber \\ 
P^{QM}(b|Y)&=& \mbox{Tr}(I \otimes y_b \ \rho_{AB}) \ .
\end{eqnarray}

If the detectors aren't perfect, i.e. $\eta<1$, a supplementary outcome is 
possible, corresponding to the case where the detector don't fire. We 
denote this outcome by the symbol $\emptyset$. We then have the 
probabilities: \begin{eqnarray}\label{qceta} P^{QM}_{ \eta}(a,b|X,Y)&=&
\eta^2 \ P^{QM}(a,b|X,Y) \quad a,b\neq 
\emptyset \ ,
\nonumber \\ 
P^{QM}_{\eta}(\emptyset,b|X,Y)&=&
\eta(1-\eta)\  P^{QM}(b|Y) \quad b\neq
\emptyset\ ,
\nonumber \\ 
P^{QM}_{\eta}(a,\emptyset|X,Y)&=&
\eta(1-\eta)\ P^{QM}(a|X) \quad a\neq 
\emptyset \ ,
\nonumber \\ 
P^{QM}_{\eta}(\emptyset,\emptyset|X,Y)&=&
(1-\eta)^2  \ .
\end{eqnarray}

In a local hidden variable theory, the quantum correlations 
(\ref{qc}) or (\ref{qceta}) are reproduced with the help of a random variable 
$\lambda$ shared by both parties. Moreover the outcomes of measurements 
performed by one of the parties are determined by the settings of the
measurement apparatus of that party only. Correlations predicted by these 
theories are thus of the form: \begin{equation}\label{lhv}
P^{LHV}(a,b|X,Y) = \int \mbox{d} \lambda \ p(\lambda) P(a|X,\lambda)
P(b|Y,\lambda)
\end{equation}
where $\lambda$ is the shared randomness. 
In the case of inefficient detectors, 
$a$ and $b$ can take either a value different from $\emptyset$, or 
the value $\emptyset$. In the latter case the LHV 
model just instructs the detectors not to fire.

\section{A LHV model that depend only on the number of settings}\label{lhvsett}

A classical theory can reproduce all the results of quantum mechanics 
if information on which measurement has been selected can flow from one side 
to the other. It is to guarantee that such mechanism cannot account of the observed 
data that measurements in Bell tests must be carried out at spatially separated 
regions. A LHV model can nevertheless exploit the limited detection efficiency 
by guessing \emph{a priori} which measurement will be performed on one side. If 
the actual measurement and the guessed one coincide, the model will output 
results in agreement with quantum mechanics. If they don't, it simply tells 
the detectors not to fire. Building a LHV 
model out of this idea will enable us to prove the following bound:

{\bf Theorem 1 :} In experiments where Alice can choose between $M_A$ 
measurements and Bob $M_B$, the maximum detection efficiency $\eta^*$ 
for which a LHV model exist is at least
\begin{equation}\label{bound1}
\eta^*\geq\frac{M_A+M_B-2}{M_AM_B-1} \ .
\end{equation}

{\bf Proof :} The proof consist of constructing a LHV model that reproduce 
the correlations (\ref{qceta}) with $\eta$ given by the bound. In 
this model, the local hidden variable $\lambda$ consist of the pair 
$\lambda=(a',X')$ where $X'$ correspond to one of the $M_A$ possible 
measurements of Alice and $a'$ to one of the possible outcomes. $X'$ is chosen 
with probability $1/M_A$ and $a'$ with probability $P^{QM}(a'|X')$, so that 
$p(\lambda)=P^{QM}(a'|X')/M_A$. If Alice's actual measurement $X$ coincides 
with 
$X'$ (this occurs with probability $1/M_A$), Alice outputs $a'$,
otherwise she outputs $\emptyset$. We thus have 
$P(a|X,\lambda)=\delta_{aa'}\delta_{XX'}$ if $a\neq \emptyset$ and 
$P(a|X,\lambda)=1-\delta_{XX'}$ if $a=\emptyset$. On the other hand, Bob 
always gives an output different from $\emptyset$. 
He randomly chooses a result $b$ using the
probability distribution   
$P(b|Y,\lambda)=P^{QM}(a',b|X',Y)/P^{QM}(a'|X')$. 

So far, Alice's efficiency $\eta_A$ is equal to $1/M_A$ and Bob's efficiency $\eta_B=1$. 
To make the protocol symmetric, Alice and Bob must exchange their role part of 
the time. This is done with the help of a supplementary hidden variable which 
tells both parties to run the protocol as above with probability $p$ and the
permuted one with probability $1-p$. There is then one problem left with the 
model, it never happens that both detector don't fire. This can be
corrected by  
adding yet another supplementary LHV that instruct Alice's and Bob's
detectors to both 
produce the result $\emptyset$ with probability $(1-q)$ and to proceed as 
above with probability $q$.  Using (\ref{lhv}), it is then not difficult to 
check that our model produces the following correlations: 
\begin{eqnarray} 
P^{LHV}(a,b|X,Y)&=& q\left(\frac{p}{M_A}+\frac{1-p}{M_B}\right)\ 
P^{QM}(a,b|X,Y)\ ,
\nonumber \\ P^{LHV}(\emptyset,b|X,Y)&=& q\ p\frac{M_A-1}{M_A} \  
P^{QM}(b|Y)\ ,
 \nonumber \\ P^{LHV}(a,\emptyset|X,Y)&=& q 
\left(1-p\right)\left(\frac{M_B-1}{M_B}\right) \ P^{QM}(a|X)\ ,
 \nonumber \\ 
P^{LHV}(\emptyset,\emptyset|X,Y)&=& 1-q  \ . \end{eqnarray}
These correlations are similar to the quantum ones (\ref{qceta}), modulo the 
detection probabilities, i.e. the probability that Alice's and Bob's, Alice's
only, Bob's only or neither detector fire. The two distributions will be 
identical if these detection probabilities coincide:
\begin{eqnarray}\label{q=lhv1}
 \eta^2&=& 
p\left(\frac{q}{M_A}+\frac{1-q}{M_B}\right) \ ,\nonumber \\ 
\eta(1-\eta)&=& p\ 
q\frac{M_A-1}{M_A} \ ,
\nonumber \\ \eta(1-\eta)&=& p 
\left(1-q\right)\left(\frac{M_B-1}{M_B}\right) \ ,
\nonumber \\ (1-\eta)^2&=& 1-p 
\ . \end{eqnarray}
Solving for $\eta$ gives the right-hand side of (\ref{bound1}). $\Box$

When $M_A=M_B=2$, the simplest non-trivial case, our bound predicts 
$\eta^*\geq 2/3$. It follows from Eberhard's result \cite{Eb} that this value is 
optimal. Indeed Eberhard has shown that there exists a 2-settings Bell 
experiments performed on a non-maximally entangled state of two qubits that
violate local realism for value of $\eta$ arbitrarily close to $2/3$. For larger
values of $M_A$ and $M_B$, $\eta^*$ as given by (\ref{bound1}) decreases and
tends to zero when both $M_A$ and $M_B$ tends to infinity. It is not known
whether our bound can be attained by quantum mechanics in these situations.
However note that there are quantum correlations produced by experiments with
exponentially many measurement settings, and for which $\eta^*$ is exponentially
small \cite{M}. It is thus at least possible to approach the bound
(\ref{bound1}) for large $M_A$, $M_B$.

We have attempted to generalise this result to the case of many
parties. For simplicity we have considered the case 
where each party can choose
between the same number $M$ of measurements.

We have only been able to prove our strongest result for less than 500 
parties because we had to resort to numerical computations to finish the proof. 
We state it as a conjecture:

{\bf Conjecture 2 (proven for $N \leq 500$):} 
In a Bell experiment with $N$ parties, each of
whose measuring apparatus can have $M$ settings,
\begin{equation}\label{bound2'}
\eta^* \geq \frac{N}{(N-1)M+1} \ .
\end{equation}

When the number of measurements on each site is $M=2$, the
bound (\ref{bound2'}) reduces to
\begin{equation}\label{m=2}
\eta^*\geq \frac{N}{2N-1}
\end{equation}
For two parties, we recover Eberhard threshold $\eta^*\geq 2/3$ and as
we have already mentioned this 
bound can be saturated by quantum mechanics. However, the threshold 
(\ref{m=2}) 
can be saturated by quantum mechanics for the other values of $N$ as 
well. Indeed Larsson and Semitecolos \cite{Larsson2} have generalized Eberhard's 
result to the case of many parties and have shown that $N$ qubits in a 
non-maximally entangled  
state can lead to violation of local realism for detection efficiencies $\eta$ 
arbitrarily close to (\ref{m=2}) for any $N$.

For number of measurements settings $M>2$, it is not known whether the bound
(\ref{bound2'}) can be saturated. However one can come close to
saturating it when the number of parties is large.  Indeed 
for large $N$, fixed $M$, eq. \ref{bound2'} becomes $\eta^* \geq 1/M +
O(1/N)$. 
And in
\cite{buhr} it is shown that there 
exists a measurement scenario for $M=2^l$ ($l=1,2,\ldots$) 
settings performed on 
$N$ qubits that exhibit non-locality for value of $\eta$ approaching $1/M$ as 
$N\rightarrow \infty$ for fixed $l$.

As a final remark, note that our conjecture 
seems quite constraining as regards the 
possible decrease of $\eta^*$ by increasing the number of parties. Indeed, for
fixed $M$, replacing $N=2$ by $N\rightarrow \infty$ one can expect at best a 
decrease of $\eta^*$ by a factor of $2M/(M+1)\leq 2$. From the resistance to 
detection inefficiency point of view, it seems thus more advantageous to 
consider experiments with many settings than with many parties.

As mentioned above we have not been able to prove eq. (\ref{bound2'})
for all numbers of parties. However we have been able to prove a
weaker result valid for any number $N$ of parties. In this
weaker result we do not ask the LHV model to reproduce all the quantum
correlations. Rather we only ask that if {\em all the detectors
click}, then the correlations exactly coincide with the quantum
correlations. On the other hand we do not put any constraint on the
correlations when one or more of the detectors do not click. This type
of model has been considered previously
in \cite{M,buhr}.

{\bf Theorem 3 : } Consider Bell experiments with $N$ parties and $M$
measurements settings per site. We require that if all detectors
click, the correlations should coincide with the quantum correlations,
but we do not put any condition on the correlations when one or more
of the detectors do not click. Then 
the maximum detection efficiency $\eta^*$ for which a
LHV model exists satisfies
\begin{equation}\label{bound2}
\eta^*\geq \frac{1}{M^{(N-1)/N}}
\end{equation}

We begin by proving Theorem 3. We then turn to the arguments behind
Conjecture 2.

{\bf Proof of Theorem 3 :}

As in Theorem 1, we can build a LHV model to reproduce the
correlations based on the remark that it is possible to predict outcomes
for all measurements performed at one site if measurements are guessed at the
other sites. A LHV will thus predetermine particular measurements and
corresponding outcomes for $N-1$ of the parties. If the guessed and
the actual measurements coincide which happens with probability $1/M$, these
parties output the selected result, if not, which happens with probability
$(M-1)/M$ their detectors keep quiet. Assuming that the measurements performed
by the other parties are the ones specified by the hidden variable, the last
party always output a result different from $\emptyset$. Since each party has
the choice between the same number $M$ of measurements there is no privileged
site and each party has the same probability $1/N$ to be selected as the
special one for which the detector always fire.

Thus when all detectors click, which occurs with probability $1/M^{(N-1)}$, 
the results obtained
will agree with those of quantum mechanics. This probability should be
identified with $\eta^N$, the probability that all detectors
click. This proves Theorem 3. $\Box$

We now turn to Conjecture 2.

{\bf Proof of Conjecture 2 for $N  \leq 500$ :}

The basic idea is to try to use the LHV model introduced in the
proof of Theorem 3  to reproduce all the correlations, 
and not only the restricted one obtained when all detectors clicks.

Note that in the model introduced in the
proof of Theorem 3,  
a detector clicks only if we are sure that it will output an
answer that agrees with quantum mechanics. The only way for the LHV
model and quantum mechanics to differ is thus in the probabilities
that the detectors click, not in the correlations of outputs
\emph{conditional} on the firing of the detector. Similarly to (\ref{q=lhv1}),
predictions of quantum mechanics and the LHV model will therefore be
identical provided they give the same detection probabilities
$q(k)$ that $k$ given detectors don't
fire and the remaining $N-k$ do. For quantum mechanics these
probabilities are given by \begin{equation}\label{qkqm}
q^{QM}(k)=\eta^{N-k}(1-\eta)^{k} \end{equation}
In particular this implies that the ratios
\begin{equation}\label{ratios}
\frac{q^{QM}(k)}{q^{QM}(k+1)}=\frac{\eta}{1-\eta}
\end{equation}
are independent of $k$.

The LHV model introduced in Theorem 3 predicts the probabilities
\begin{equation}\label{q0k}
q^{LHV}(k)=\frac{N-k}{N}\frac{(M-1)^{k}}{M^{N-1}}
\end{equation}
(see eq. (\ref{qik}) with $i=0$ and the explanation in the paragraph following 
eq. (\ref{qik})).
It has thus the property 
that \begin{equation}\label{qoverq}
{q^{LHV}(0) \over q^{LHV}(1)} = { N \over (N-1) (M-1)}
\end{equation}
Using eq. (\ref{ratios}) and solving for $\eta$ yields
eq. (\ref{bound2'}). This is the basis for Conjecture 2.

But from (\ref{q0k}) we also deduce
\begin{equation}
\frac{q^{LHV}(1)}{q^{LHV}(2)}>\frac{q^{LHV}(0)}{q^{LHV}{1}}
\end{equation}
in contradiction with (\ref{ratios}). Furthermore the model
introduced in Theorem 3
never instructs the $N$ dectector to keep quiet
simultanously.

We can try to correct the model so as to recover eq. (\ref{ratios}, while 
leaving (\ref{qoverq}) unchanged, by increasing the probability $q^{LHV}(k)$, 
$k\geq 2$ that more than one  party does not fire. 

A natural way to extend our protocol so that it can reproduce
the whole set of correlations is to introduce the
possibility for 
it to constrain $i$ ($i=2,\ldots, N$) of the parties to output $\emptyset$,
similarly to the proof of Theorem 1 where part of the time Alice and Bob had
both to produce result $\emptyset$

The new LHV model will therefore be build out of a family of $N$ protocols
${\cal P}_i$ ($i=0,2,\ldots, N)$. In protocol ${\cal P}_i$, a subset of $i$ of
the $N$ parties is forced to output 
$\emptyset$ independently of the measurement
performed at these $i$ sites. Since they are ${N \choose i}$ possible
choices 
of
$i$ parties among the $N$, the probability that one particular subset is chosen
is $1/{N \choose i}$. The protocol then works as before with $N$ replaced by
$N-i$. The probabilities $q^i(k)$ that $k$ given detectors don't fire and the
remaining $N-k$ do for protocol ${\cal P}_i$ are given by
\begin{equation} \label{qik} q^{i}(k)=\begin{cases} 0 &
\text{$k<i$,}\\ \displaystyle \frac{{k\choose i}}{{N \choose
i}}\frac{N-k}{N-i}\frac{(M-1)^{k-i}}{M^{N-i-1}} &\text{$k\geq i$} \\ 1 &
\text{$k\mbox{ and } i=N$.} \end{cases} \end{equation}
The first and the last case of (\ref{qik}) are trivial. Indeed, in our
protocols at least $i$ parties produce the result $\emptyset$ so that their
contribution to events where $k<i$ parties don't fire is null. On the other
hand, the protocol ${\cal P}_N$ always output $\emptyset$ for the $N$
parties. For the remaining case when $k\geq i$ detectors don't click, the
subset of $i$ parties that are forced to output $\emptyset$ must certainly be
included in the subset of the $k$ parties that don't click. Since they are ${k
\choose i}$ subset out of the ${N \choose i}$ possible that satisfy this
condition we have the term ${k \choose i}/{N \choose i}$. Secondly, the
special party for which the detector always fire can not be one of the $k$ not
clicking. There thus remain only $N-k$ possibilities over the $N-i$
original one, hence the term $(N-k)/(N-i)$. Finally, in the remaining $N-i-1$
parties $k-i$ of them must output $\emptyset$, which happens with probability
$(M-1)^{k-i}/M^{N-i-1}$.

If the LHV model instructs 
to use protocol ${\cal P}_i$ ($i=0,2,\ldots N$) with 
probability $p_i$ we find
\begin{eqnarray}
q^{LHV}(k)&=&p_0q^0(k)+\sum_{i=2}^Np_iq^i(k) \nonumber \\
&=&p_0q^0(k)+\sum_{i=2}^kp_iq^i(k)
\end{eqnarray}
since $q^i(k)=0$ for $i>k$.

As already stated above our model predict the correct probabilities
\emph{conditional} on the firing of the detectors. It will thus
properly reproduce the quantum probabilities obtained in an experiment 
provided the detection probabilities satisfy $q^{LHV}(k)=q^{QM}(k)$ or
\begin{equation}\label{qm=lhv}
\eta^{N-k}(1-\eta)^{k}=p_0\ q^0(k)+\sum_{i=2}^k p_i \ q^i(k) \quad \mbox{for all
$k$}. \end{equation}
This will be the case if this set of equations for the $p_i$
admits a solution such that the $p_i$ are positive and sum to one, i.e. they
form an actual probability distribution.

The fact that 
they sum to one is already implied by the structure of (\ref{qm=lhv}).
Indeed summing both sides of (\ref{qm=lhv}) over all possible
subset of parties for which the detector fire and don't fire, we
deduce that $p_0 +\sum_{i=2}^N p_i=1$, since $\sum_k {N
\choose k} \eta^{N-k}(1-\eta)^k=\sum_k {N \choose k}q^i(k)=1$ .

To check wether the $p_i$ are positive we use
\begin{equation}
\frac{\eta}{1-\eta}=\frac{q^0(0)}{q^0(1)}=\frac{\sum_{i=0}^{k-1}p_iq^i(k-1)}{
\sum_{i=0}^{k}p_iq^i(k)}
\end{equation}
to write
\begin{equation}
p_k=\frac{1}{q^k(k)}\sum_{i=0}^{k-1}p_i\left(\frac{q^0(1)}{q^0(0)}q^i(k-1)-q^i(
k)\right)
\end{equation}
This define recursively the $p_i$ starting from $p_0$=cst$>0$ and $p_1=0$. Note
that the $p_i$ depends of $N$ and $M$. If we define $r_k=M^k/(M-1)^k\ p_k$ we
obtain for the $r_k$ the recursive definition
\begin{equation}\label{rk}
r_k=\frac{1}{q'^k(k)}\sum_{i=0}^{k-1}r_i\left(\frac{q'^0(1)}{q'^0(0)}q'^i(k-1)-
q'^ i ( k)\right)
\end{equation}
where $q'^i(k)=M^{N-i-1}/(M-1)^{k-i}\ q^i(k)$. Since the $q'^i(k)$ are indepent 
of $M$ so are the $r_k$. If all the $r_i$ are positive for 
given $N$ it thus follows that all the $p_i$ are also positive for that given 
$N$ and for all values of $M$. We checked this positivity condition for the 
$r_i$ for $N\leq 500$ using a symbolic mathematics software (Mathematica) 
that performs exact computations (indeed, non-linear recursive equations as 
(\ref{rk}) are sensitive to small numerical perturbations and we didn't find any 
stable method of solving (\ref{rk}) using finite precision arithmetics). This 
concludes the proof of Conjecture 2 for $N \leq 500$. $\Box$

\section{A LHV model that approximately reproduces the quantum correlations for
given dimensionality}\label{lhvdim}

We now present a LHV model inspired by the communication protocol described
in \cite{MBCC}. Though this model is probably not optimal, it shows that it is
in principle possible to build LHV models that depend only of the dimension of
the quantum system. In this model $\eta$ decreases exponentially with $d$. This
behavior of $\eta$ must be shared by all models that depends only on the
dimension since in \cite{M} it is shown that there are quantum correlations
which are non local even when the detector efficiency is exponentially small in
$d$. Note however that the quantum correlations in \cite{M} require an almost 
complete absence of noise to exhibit non-locality, whereas the model described 
below reproduces noisy correlations (although the amount of noise decreases with 
the dimension for fixed $\eta$).

Note: for simplicity of notation, in this section all the probabilities 
$P^{LHV}$ or $P^{QM}$ we compute or refer to are probabilities conditional on 
the firing of both the detectors.

{\bf Theorem 4 :} For measurements performed on
the maximally entangled state
$|\Phi\rangle=\sum_{i=0}^{d-1}\frac{1}{\sqrt{d}}|ii\rangle$ and for given $\epsilon <
2d$, there exists a LHV model that produces a probability distribution
$P^{LHV}(a,b|X,Y)$ such that for all $X,Y$, $P^{LHV}(a|X)
= P^{QM}(a|X)$, $P^{LHV}(b|Y) = P^{QM}(b|Y)$ and $|P^{LHV}(a,b|X,Y) -
P^{QM}(a,b|X,Y)| \leq  \epsilon P^{QM}(a|X)P^{QM}(b|Y) $ when the efficiency 
of the detectors is 
\begin{equation} \eta
=\left({\epsilon \over 4d}\right)^{2(d-1)}
\end{equation}

{\bf Proof :}
We recall that Alice and Bob carry out the POVM's $X$ and $Y$ with elements
$x_a$ and $y_b$. Without loss of generality we can suppose that $x_a$ and $y_b$
are rank one \cite{bar}. We rewrite them as
$$x_a = |x_a| \ |x_a\rangle\langle x_a |\quad , \quad
y_b = |y_b| \ |y_b\rangle\langle y_b |$$
where $|x_a\rangle , |y_b\rangle$ are normalized states.
In the case of the maximally entangled state, the marginals and the joint
outcome probability are
\begin{eqnarray}
\begin{array}{lcr} P^{QM}(a|X)=\frac{|x_a|}{d}& \mbox{,}&
P^{QM}(b|Y)=\frac{|y_b|}{d} \end{array} \nonumber \\
P^{QM}(a,b|X,Y) = \frac{1}{d}|x_a| |y_b| |\langle x_a^* | y_b\rangle|^2
\end{eqnarray}
where $|x_a^* \rangle=\sum_i x_a^{i*}|i\rangle$ with $x_a^i$ the components of
$|x_a\rangle$ in the basis where $|\Phi\rangle=\frac{1}{\sqrt{d}}\sum_i
|ii\rangle$.

The local hidden variable  consists of the classical description
of a pure quantum state $|\phi\rangle$. This state is uniformly chosen
in the Hilbert space using the invariant measure over $SU(d)$.
Alice's strategy is the following: she first chooses $a$ with
probability $|x_a|/d$, in agreement with the marginal probability
$P^{QM}(a|X)$. Having fixed $a$ she then computes  $s=|\langle \phi|
x_a\rangle|^2$. If $s<\cos^2\delta$, she outputs ``no result''. If $s \geq
\cos^2\delta$, she outputs $a$ (where $\delta >0$ will be fixed below). The
probability $Q$ for Alice to give an outcome is
\begin{equation}
Q =
\int_{SU(d)} \mathrm{d} \phi \
\Theta (  |\langle \phi | x_a\rangle|^2  -\cos^2\delta)
\end{equation}
To compute this expression we write
$|\phi\rangle=\cos \theta|x_a\rangle+e^{i\rho}\sin \theta|\phi_{d-1}\rangle$
where $|\phi_{d-1}\rangle$ lies in the subspace orthogonal to $|x_a\rangle$.
Since
$\mathrm{d}\phi=\frac{d-1}{\pi}\cos\theta(\sin\theta)^{2d-3}\ \mathrm{d}\theta
\ \mathrm{d}\rho\ \mathrm{d}\phi_{d-1}$ we find
\begin{eqnarray}
Q &=& 2(d-1)\int_0^{\pi/2} \mathrm{d}
\theta \cos\theta (\sin\theta)^{2d-3} \Theta (\cos^2\theta -
\cos^2\delta) \nonumber\\
&=&(\sin\delta)^{2(d-1)} \ .
\end{eqnarray}
As expected, the probability to give an outcome is independent of Alice's 
particular result $a$.
 
Bob's strategy is as follows: he gives output $b$ with 
probability \begin{equation}
P(b|Y,\phi) = |y_b| | \langle\phi^* | y_b\rangle|^2 \ .
\end{equation}
This results in the marginal probability
\begin{eqnarray}
P^{LHV}(b|Y)&=&\int_{SU(d)} \mathrm{d} \phi P(b|Y,\phi) \nonumber \\
&=&2(d-1)\ |y_b|\int_0^{\pi/2}\mathrm{d}\theta \cos^3\theta (\sin\theta)^{2d-3}
\nonumber\\ &=&\frac{|y_b|}{d}
\end{eqnarray}
where we have taken
$|\phi\rangle=\cos\theta|y_b^*\rangle+e^{i\rho}\sin\theta|\phi_{d-1}\rangle$ 
and $|\phi^*_{d-1}\rangle $ orthogonal to $|y_b\rangle$ to pass from the first 
line to the second one.

Let us now compute the joint probability of outcomes $a$ and $b$ given that an
outcome has been produced. This is
\begin{eqnarray}
\lefteqn{P^{LHV}(a,b|X,Y)}\nonumber\\
 &=& \frac{1}{Q}\int_{SU(d)} \mathrm{d}\phi
\ P(a|X,\phi)P(b|Y,\phi)\nonumber \\
&=& \frac{1}{Q}\int_{SU(d)} \mathrm{d}\phi \
\frac{|x_a|}{d} \Theta (  |\langle \phi | x_a\rangle|^2  -\cos^2\delta) |y_b|
|\langle \phi^* | y_b\rangle|^2 \nonumber\\
\end{eqnarray}
To compute how much this differs from the true probability, let us
evaluate
\begin{eqnarray}
D =  |\langle \phi^* |y_b\rangle|^2 - |\langle x_a^* |y_b\rangle|^2 \ .
\end{eqnarray}
Writing
$|\phi\rangle=\cos\theta|x_a\rangle+e^{i\rho}\sin\theta|\phi_{d-1}\rangle$
where $\langle x_a|\phi_{d-1}\rangle=0$ we find
\begin{eqnarray}
|D| &=& |\  -\sin^2\theta |\langle x_a^* |
y_b\rangle|^2  + \sin^2\theta |\langle \phi_{d-1}^* |
y_b\rangle|^2 \nonumber \\
&&+ (\sin\theta\cos\theta \langle x_a^* |
y_b\rangle \langle y_b | \phi_{d-1}^*\rangle + c.c).\ |
\nonumber\\
&\leq&  \sin^2\theta + 2 \sin\theta \ .
\end{eqnarray}
From which we deduce
\begin{eqnarray}
\lefteqn{|P^{LHV}(a,b|X,Y)-P^{QM}(a,b|X,Y)|} \nonumber \\
&\leq&\frac{1}{Q}\frac{|x_a|}{d}|y_b| 2(d-1)
\int_0^{\pi/2} \mathrm{d}\theta \cos\theta(\sin\theta)^{2d-3} \nonumber \\
&& \times \Theta
(\cos^2\theta-\cos^2\delta) (\sin^2\theta+2\sin\theta)  \nonumber \\
&\leq& \frac{1}{Q}\frac{|x_a|}{d}  |y_b|
2(d-1) (\sin^2\delta+2\sin\delta)\nonumber \\
&&\times  \int_0^{\delta}
\mathrm{d}\theta \cos(\theta)(\sin\theta)^{2d-3} \nonumber \\
&=&\frac{1}{d}|x_a||y_b| (\sin^2\delta+2\sin\delta) \nonumber
=\epsilon P(a|X)P(b|Y) \end{eqnarray}
where we have taken $\epsilon=d(\sin^2\delta+2\sin\delta)$.

In the above protocol the roles of Alice and Bob are not
symmetric and it never happens 
that both detectors don't fire. Upon letting them
take randomly one of the two roles above and forcing both detectors to stay
quiet part of the times, as in the previous LHV models, one sees that the model
we have constructed has detector efficiency $\eta/(1-\eta) =2
Q/(1-Q)$ or
\begin{eqnarray}
\eta&=&\frac{2 (\sin\delta)^{2(d-1)}}{1+(\sin\delta)^{2(d-1)}}\nonumber \\
&\geq&\sin\delta^{2(d-1)} \nonumber\\
&\geq& \left(\frac{\epsilon}{4d}\right)^{2(d-1)}
\end{eqnarray}
since $\sin\delta\geq \epsilon/2d-\epsilon^2/8d^2\geq \epsilon/4d$ when
($\epsilon <2d$). $\Box$

\section{Conclusion}\label{Conclusion}
We have exhibited LHV models that depend only on the 
dimensionality of the quantum system or only on the number of settings of each 
party's measurement apparatus. These models show that there exist general 
constraints on the violation of local realism independently of the particular 
settings of Bell experiments. They help point out which parameters are 
important when trying to find quantum experiments that exhibit strong non 
locality. For instance the existence of these LHV models served as a guiding 
principle for a recent numerical search that yielded several Bell inequalities 
resistant to detection inefficiency \cite{MPR}.

Our models can also have implications in the design of loophole-free tests of 
Bell inequalities. Loophole-free tests of Bell inequalities are important 
both from a fondamental point of view and for the security of some quantum 
cryptographic protocols \cite{Larsson3}. In experiments involving photons the 
detection loophole remain the last serious loophole to be closed. It would 
therefore be interesting to find Bell scenarios that violate local realism for 
efficiencies of the detectors close to the actual value of our current 
photo-detectors. Our result shows that to go beyond Eberhard's threshold of 
$\eta^*\geq 2/3$ (or to go beyond Larsson and Semitecolos's threshold for many 
parties) it is necessary to consider Bell experiments with more than two 
measurements per site. This strengthens the recent interest in Bell inequalities 
involving many measurements settings which have been shown to exhibit more 
strongly the non-locality of quantum mechanics than usual Bell inequalities 
based on two settings \cite{MPR,kazli,zuko}.

\begin{acknowledgments}
We acknowledge support by the EU fifth framework projects EQUIP, 
IST-1999-11053, and RESQ, IST-2001-37559. S.M. 
is a research associate of the Belgian
National Research Foundation.
\end{acknowledgments}

\end{document}